\documentclass[a4paper]{jpconf}
\usepackage{graphicx}

\begin{document}

\title{The Galactic Center Magnetosphere}

\author{Mark Morris}

\address{Department of Physics \& Astronomy, University of California, Los Angeles, CA 90095-1547, USA}

\ead{morris@astro.ucla.edu}

\begin{abstract}
The magnetic field within a few hundred parsecs of the center of the Galaxy is an essential component of any description of that region.  The field has several pronounced observational manifestations: 1) morphological structures such as nonthermal radio filaments (NTFs) -- magnetic flux tubes illuminated by synchrotron emission from relativistic electrons -- and a remarkable, large-scale, helically wound structure, 2) relatively strong polarization of thermal dust emission from molecular clouds, presumably resulting from magnetic alignment of the rotating dust grains, and 3) synchrotron emission from cosmic rays.   Because most of the NTFs are roughly perpendicular to the Galactic plane, the implied large-scale geometry of the magnetic field is dipolar.  Estimates of the mean field strength vary from tens of microgauss to $\sim$ a milligauss.  The merits and weaknesses of the various estimations are discussed here.  If the field strength is comparable to a milligauss, then the magnetic field is able to exert a strong influence on the dynamics of molecular clouds, on the collimation of a Galactic wind, and on the lifetimes and bulk motions of relativistic particles.  Related to the question of field strength is the question of whether the field is pervasive throughout the central zone of the Galaxy, or whether its manifestations are predominantly localized phenomena.  Current evidence favors the pervasive model.  

\end{abstract}

\section{Introduction}

The magnetic field at the center of the Galaxy (hereafter, the "field") has been studied with a wide variety of techniques for over 20 years, and while there is some consensus that the predominant, global geometry within the central 200 - 300 parsecs is poloidal, the discussion at this workshop has emphasized that there is no universal agreement on the strength of the field and on the extent to which the field strength varies from one place to another.  In this review, I summarize the evidence characterizing the various points of view.  Earlier reviews of the Galactic center magnetic field have described many of the central points that have been known for some time \cite{M1,M2, M3, M4, MS96, Novak05}, but recent observations have added considerably to the information that can be brought to bear on this discussion. 

The primary probe of the large-scale field has been radio observations of polarized, filamentary structures which, while typically $<$ 0.5 pc in width, are tens of parsecs in length.  The strong radio polarization, and the occasional filamentary counterpart at X-ray wavelengths \cite{YZ_Xfil05} indicate that the emission is synchrotron radiation, and the position angle of the polarization, once corrected for Faraday rotation, confirms that the magnetic field lies along the filaments \cite{Tsuboi86, YZWP97, Lang99, Langhere}. The almost invariant curvature of the filaments, and their absence of distortion in spite of clear interactions with the highly turbulent interstellar medium, led Yusef-Zadeh \& Morris (1987 \cite{YZM87}, see also \cite{MS96}) to note that the implied rigidity of the filaments requires a field strength on the order of a milligauss, which is surprisingly large, given the scale of these structures. 

The orientation of the most prominent NTFs is roughly perpendicular to the Galactic plane, as illustrated in Figure 1, a schematic diagram depicting all filaments identified in the $\lambda$20-cm VLA survey by Yusef-Zadeh et al.\ (2004 \cite{YZHC04}).  Because the individual filaments define the local field direction, the ensemble of filaments has been interpreted in terms of a predominantly dipolar field, extending at least 200 pc along the Galactic plane \cite{Nord04}.  The deviations from perfect verticality of many of the filaments can be ascribed to a global divergence of the field above and below the Galactic plane.  The short, nonconforming filaments are discussed in $\S$2.3 (and \cite{LaRosahere}).

\begin{figure}[h]
\includegraphics[width=36pc]{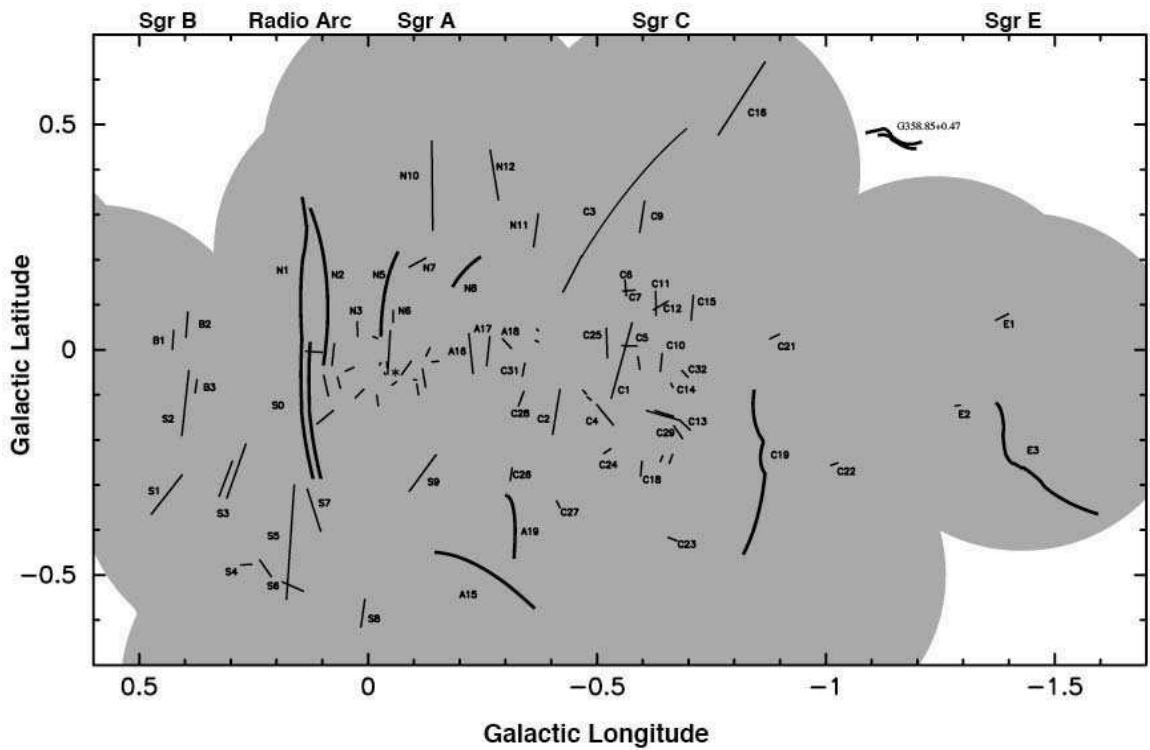}
\caption{\label{label}Schematic map showing the radio filaments catalogued by Yusef-Zadeh et al. (2004, \cite{YZHC04}) in the course of their $\lambda$20-cm survey of the Galactic center.}
\end{figure}

Quite a different probe of the magnetic field is provided by mid- and far-IR observations of thermal dust emission from magnetically aligned dust grains.  The rotation axes of dust grains align with the magnetic field by dissipative torques \cite{Hildebrand88}, leading to a net polarization of the thermal emission such that the E-vector is perpendicular to the magnetic field.  This probe, however, is strongly dominated by dense, warm clouds, so it is quite different from the NTFs, which sample the field in the intercloud medium occupying most of the volume of the Galactic center.  The magnetic field implied by the polarized dust emission is parallel to the Galactic plane \cite{Werner88, Morris92, Hildebrand93, Novak03, Chuss03}, and thus perpendicular to the large-scale intercloud field revealed by the NTFs.  The perhaps surprising orthogonality of these two systems can be understood in terms of the tidal shear suffered by molecular clouds inhabiting the central molecular zone (CMZ).  Any portion of a molecular cloud located a distance R$_{gc}$ pc from the Galactic center, and having a density less than 10$^4$ cm$^{-3}$ [75 pc/R$_{gc}]^{1.8}$ is subject to such shear \cite{Gusten89, Stark04}, so cloud envelopes tend to get stretched into tidal streams that may subtend a large angle at the Galactic center (e.g., \cite{Tsuboi99}).  Any magnetic field within the clouds -- presumably flux-frozen to the partially ionized molecular gas -- will thus be deformed into an azimuthal configuration, with the field lines oriented predominantly along the direction of the shear \cite{Werner88}.   

There is little evidence that the cloud and inter-cloud environments are magnetically coupled to each other in any significant way, as might have been expected if the field lines were anchored to the cloud layer, and if the rotation of the cloud layer thus imposes a global twist upon the vertical field \cite{Uchida85, Shibata87}.  The most prominent NTFs show very little deformation where they pass through the Galactic plane and interact with gas in the CMZ ({\it e.g.}, \cite{YZM87}).  Some case can be made that Faraday rotation measurements are consistent with the geometry of a twisted, large-scale field (\cite{Novak05}, and references therein), but these data remain too sparse to draw any firm conclusions.  

If, as the evidence does indicate, the magnetic field is not anchored in the CMZ, then it is either anchored in the essentially non-rotating Galactic halo or beyond, or it arcs back to the Galactic plane at relatively large radii and is anchored there.  In either case, the field lines do not rotate with the CMZ, and the molecular clouds move through the field with a large relative velocity.  This gives rise to an induced {\bf v}$\times${\bf B} electric field at cloud surfaces (10$^{-4}$ B(mG) V/cm) which can accelerate particles, drive currents and contribute to the cloud heating \cite{Benford88, Morris89}.  

The residence time of clouds in the Galactic center is a few hundred million years as a result of angular momentum loss resulting from both dynamical friction and magnetic drag \cite{Stark91, M2, Belmont06}, so it is not clear how clouds forming at the outside edge of the CMZ 
\cite{Binney91} will retain any magnetic contact with their surroundings as they migrate inwards through the vertical field.  Any original connection between the cloud and extra-cloud fields could have pinched off during the inward migration, leaving the clouds magnetically isolated.  If typical cloud lifetimes are less than the inspiral times of clouds, presumably because clouds are sheared in the tidal field, then the situation is more complex, but these comments can still apply to sheared cloud streams and the new clouds that reform as the streams interact with each other.  

The remainder of this review focuses on several topics of current interest -- both observational and theoretical -- and culminates in a description of what I think are some of the most important open questions.

\section{Uniformity of the Galactic Center Field}

\subsection{Pressure Confinement of Magnetic Structures}
Regardless of the magnetic field strength, the pressure of the interstellar medium in the CMZ is very large compared to the Galactic disk \cite{SpergelBlitz92}.   A hot diffuse gas (T $\sim$ 10$^8$ K, n $\sim$ 0.04 cm$^{-3}$) that pervades much of the volume of the Galactic center \cite{Muno04, Belmont05,Belmont07} has a pressure of 6 x 10$^{-10}$ dynes cm$^{-2}$, and is in approximate pressure equilibrium with the warm ($\sim$150 K, low-density molecular medium \cite{Dahmen98, RodFer04}, if the velocity dispersion of $\sim$20 km s$^{-1}$ is used to calculate a turbulent pressure.  This pressure is at least two orders of magnitude higher than is characteristic of the Galactic disk.  The magnetic field, on the other hand, has a pressure of 4 x 10$^{-8}$ B(mG)$^2$ dynes cm$^{-2}$.  Consequently, if the magnetic field strength in observed magnetic field structures is $\sim$ a milligauss, then those structures are not confined, and would expand and disappear on a short time scale.  This consideration led to the argument that a milligauss magnetic field must be pervasive throughout the CMZ \cite{MYZ89}; the strong and extended magnetic field would then provide its own support.  In this view, the NTFs are then simply illuminated magnetic flux tubes into which relativistic electrons have been injected, and along which the electrons are constrained to flow \cite{M1}.  A ring current at the outer edge of the CMZ, or distributed over some range of radii there, is required to generate and confine the overall dipole field \cite{MS96}.  

\subsection{Models of Localized Magnetic Structures}  
The alternative to a strong, pervasive field is that the NTFs represent localized peaks in the magnetic field strength.
A force-free magnetic field configuration might be considered as a way of tying a local current to a local enhancement of the magnetic field strength {\cite{YZMC84, YZM87b}, but unless the overall configuration is pressure confined, it will be transient and short-lived.  

A recent suggestion by Boldyrev \& Yusef-Zadeh \cite{BoldYZ06} is that the NTF's are localized structures of milligauss field strength confined by the effective pressure of large-scale turbulence in the Galactic center.   In their model, the turbulent cells expulse the field, and concentrate it in regions between the cells.  However, while the field will indeed diffuse out of a zone of strong turbulence, the turbulence itself is generally accompanied by the generation of new field at a rate at least as fast as the rate of outward diffusion.  Consequently, while this mechanism raises the interesting possibility that the geometry of the boundary field might be different from that within the turbulent zones because of the interactions of the field emanating from the different zones, it is not obvious how this mechanism would lead to a relative enhancement of the field strength at those boundaries. 
Furthermore, the turbulence in this model must be organized in such a way that the resulting magnetic filaments are predominantly vertical.  This places a strong constraint on the overall helicity distribution of plasma motions in the Galactic center.  Numerical models that address these concerns are needed to assess this model further.  

While other models for localized structures have been proposed \cite{Bicknell01, Shore99, Dahlburg02}, they lack the generality needed to account for the population and the orientations of the filaments.

\subsection{Significance of the Short Radio Streaks?}  
One relatively recent finding that has called the notion of a pervasive, uniform field into question is a population of short radio filaments, or streaks, that occupy much of the same Galactic longitude range as the prominent NTFs \cite{Nord04, LaRosa04, LaRosahere}.   These structures are largely included in figure 1.  They differ in three ways from the long-known, prominent NTFs: 
\begin{enumerate}
\item{They are quite short, $\sim$0.1 pc.}
\item{Their surface brightness is typically about 1/4 that of the prominent NTFs.} 
\item{They appear to be more or less randomly oriented, and thus do not conform to the global verticality of the prominent NTFs.  This point has been raised as an argument against a globally ordered, dipole magnetic field.  }
\end{enumerate}
Given these pronounced differences, one could argue that the radio streaks represent a different population with a separate origin, such as localized oblique shock structures, or strong local deformations of the large-scale field as a result of some local, energetic disturbance.  It is premature to conclude that they are inconsistent with a predominantly ordered, large-scale dipole field.  Further study of these features is warranted to determine whether they differ systematically from the prominent NTFs in other ways as well, such as in terms of spectral index and polarization properties, and whether they are connected to other interstellar structures in the same way that the prominent NTFs are.  

\subsection{Dynamical Consequences}

As mentioned above, a pervasive, dipole field exerts a magnetic drag force on clouds moving through it, enhancing the rate at which they spiral inwards.    If sufficiently strong, the field can also collimate winds and energetic particles that emanate from the center, creating a chimney effect.  This is consistent with observations of extended columnar radio features in nearby, radio-bright galactic nuclei \cite{HvGK83, Duric88, KeelWehrle93}, although the extent to which the energetic winds in such galaxies have been collimated by the magnetic field, as opposed to the back pressure of their stratified interstellar gas layers, has not been settled.  

Recent work by Belmont {\it et al.} \cite{Belmont05} has shown that at least the hydrogen in the hot, diffuse gas at the Galactic center is unbound, so a thermal galactic wind is implied.  A dipole magnetic field can collimate this wind to an extent that depends on the field strength, so observations of the large-scale morphology of thermal X-ray emission from the hot gas will be a useful probe of both the wind and the magnetic field.  

Cosmic rays will also be confined by a pervasive, vertical field.  This has two important consequences: first, the  residence time for cosmic rays in the Galactic center will be relatively short  (a few $\times$ 10$^5$ yr) compared to that in the Galactic disk (a few $\times$ 10$^6$ yr), because the constraint that cosmic rays diffuse primarily along the field lines implies, in the Galactic center, that they diffuse directly away from the Galactic plane, whereas in the Galactic disk, they are largely trapped by the azimuthal field.  This relatively short residence time implies a much smaller cosmic ray density than one might infer from the volume rate of supernovae alone.   This is consistent with the fact that the high-energy $\gamma$-ray emission intensity across the CMZ does not have a peak comparable in its contrast to the peak in the total column density of gas \cite{Hunter97, Strong04}.
Second, the longitudinal diffusion of cosmic rays, especially electrons, would be suppressed by a pervasive vertical field. Such diffusion -- for protons -- is assumed in a recent model for the extended TeV emission observed by HESS  invoking a single source of high-energy cosmic rays \cite{HESS05, Hintonhere}; this model is probably inconsistent with the presence of a strong, pervasive, vertical field.  

\section{Comments on Arguments for a Weak Field}

\subsection{The Minimum Energy Assumption}
A number of researchers have estimated the strength of the Galactic center magnetic field using the minimum energy assumption, also referred to as "equipartition", applied to observations of synchrotron emission from relativistic particles (e.g., \cite{LaRosa05}).  This assumption can be applied to a medium in which energy exchange takes place between particles and fields on time scales much less than the energy loss times of particles or the field generation time from macroscopic particle dynamics.   This can, for example, describe environments characterized by isotropic turbulence and tangled fields, such as the hot spots in the lobes of double radio source galaxies.  However, it is quite generally inapplicable to the Galactic center, except perhaps in very local environments in which energetic events have recently occurred.  The striking large-scale order of the Galactic center magnetic field implies that its energy content is not responding in any significant way to local fluid motions or relativistic particle dynamics.  The relativistic particles are responding to the field, but the reverse is not true.  The energy content of the Galactic center field is far greater than that of the emitting particles, and thus the field strength can be much larger than the equipartition value. 

\subsection{Zeeman Measures}

The most compelling measure of field strength would be a direct measure via the Zeeman effect.  Zeeman measures have indeed been made in Galactic center clouds in lines of both H and OH \cite{Schwarz89, Killeen92, Plante95, UchidaGusten95, Marshall95}, with the result that, where any significant Zeeman signal is seen at all, it implies a field strength on the order of a milligauss or larger.  However, there are only a few places where a significant Zeeman signal has been detected.  (We do not include in these comments the Zeeman measures deduced from 1720-MHz OH masers around Sgr A East and the circumnuclear disk, which give field strengths of 3-5 mG \cite{Yusef_OHmaser99, Loranthere}, because such masers presumably arise from locally compressed gas, and may therefore not be representative of the magnetic field on large scales.)  One strong selection effect in Zeeman measures is that the extremely broad lines of Galactic center clouds make detection of the Zeeman splitting very difficult unless the field strength exceeds $\sim$1 mG.  

Two other points must be considered when interpreting Zeeman measurements: first, they apply largely to the magnetic field within clouds or at the surfaces of clouds.  As the above discussion indicates, the magnetic field geometry in clouds is not necessarily related to the large-scale intercloud field.  Second, the Zeeman effect measures only the mean line-of-sight component of the field, so if there are field reversals along the line of sight, or if the field direction changes across the radiotelescope beam, then there is significant averaging and dilution of the Zeeman signal.  In any case, even if Zeeman measures were able to provide insight into the strength of the intercloud field, the line-of-sight restriction makes it difficult to draw conclusions about a largely vertical dipole field.  Further Zeeman measurements, not only of H and OH with improved sensitivity and spatial resolution, but also of other molecules that probe denser regions, will be very important for achieving a more complete understanding of the Galactic center field.  

\subsection{Synchrotron Lifetimes}

One argument that has been raised against  a pervasive field of milligauss strength is that the synchrotron lifetime of the electrons responsible for the nonthermal radio emission is relatively short, $\sim$10$^5$ years for electrons responsible for the 330-MHz radio emission arising from the central 4$^{\circ}\times$2$^{\circ}$ diffuse nonthermal source \cite{LaRosa05}.  So the supernova rate in the CMZ (or in the Galactic and nuclear bulges above it, since not much less than half of the relativistic electrons created in a supernova will diffuse along the field lines and reach the Galactic plane) must be somewhat larger than 1 per 10$^5$ yrs if supernovae alone are to account for the uniformity of the synchrotron emission.  The rate of only Type Ia supernovae in the Galactic bulge has been estimated at 30 per 10$^5$ yrs, \cite{Schanne06}, and in the nuclear bulge (defined in \cite{SM96, LZM02}) it is about 20 per 10$^5$ yrs, so allowing also for core collapse supernovae, the particle production rate seems abundantly sufficient, even if no particles diffuse to the Galactic center from the rest of the Galaxy \cite{CowinMorris07}, and if there is no particle reacceleration process operating.  

The synchrotron lifetimes of electrons responsible for the 5-GHz radio emission from the NTFs is only $\sim$10$^4$ years, so if they diffuse along the field lines at the Alfv\'en speed, 2200 km s$^{-1}$ B(mG)/n(cm$^{-3}$)$^{1/2}$, then the net distance they can travel before losing an appreciable amount of energy is $\sim$20 pc $\times$ B(mG)/n(cm$^{-3}$)$^{1/2}$, somewhat shorter than the length of the longest filaments (60 pc).  (The Alfv\'en speed is assumed because the diffusion is usually limited by scattering of the streaming particles off of Alfv\'en waves propagating along the field lines).  So far, observations indicate that the radio spectral index has no noticeable variation along the length of the filaments ({\it e.g.}, \cite{Lang99}).   Consequently, if the relativistic electrons are produced at a specific location along them, then the synchrotron lifetime may present a problem unless the field strength is substantially less than a milligauss.  Two possible alternatives warrant consideration: first that the diffusion along the field lines is much faster than the relatively slow rate assumed here because the magnetic field is much more rigid and smooth than in most situations where the Alfv\'en speed is invoked.  Second, a reacceleration process may take place along the filaments via shocks, wave dissipation, or reconnection, in analogy with the reacceleration processes needed to account for the persistence of highly relativistic particles in extragalactic jet sources, in spite of their synchrotron and Compton losses.

\section{The Double Helix Nebula}

A potential new probe of the Galactic center magnetic field was recently revealed at 24 $\mu$m with the Spitzer Space Telescope \cite{Morris06}.  At a distance of $\sim$100 pc toward positive Galactic latitude from the Galactic center, a nebula having the form of an intertwined double helix extends over at least 50 pc, with its long axis oriented approximately perpendicular to the Galactic plane (Figure 2).  This feature was interpreted as a torsional Alfv\'en wave propagating away from the Galactic center along the magnetic field, and driven by the rotation of the circumnuclear gas disk (CND).  The few-parsec scale of the CND matches the width of the nebula, and the wavelength of the torsional wave, 19 pc, corresponds to the $\sim$10$^4$-year rotation period of the CND if the Alfv\'en speed is 10$^3$ km s$^{-1}$.  This speed, in turn, constrains the magnetic field to have a strength of 0.5 $n^{1/2}$ mG in the context of this hypothesis, where $n$ is the hydrogen density in the medium through which the wave propagates.  The density is not known, but for values of the magnetic field ranging from 0.1 to 1 mG, a plausible density is found: $n$ = 0.04 - 4 cm$^{-3}$.
The presence of two strands has been attributed to an apparent "dumbbell" asymmetry of the driving disk (see \cite{Morris06}); the magnetic field threading the disk is concentrated into two diametrically opposed density maxima.  

A potential weakness of the torsional wave hypothesis is that the wave cannot yet be followed all the way down to its hypothetical source, the CND.  However, this also raises the question of why the double helix is visible in the first place; its mid-infrared emission is most likely thermal emission from dust, so the visibility of the nebula at its present location presumably requires that the wave has levitated charged dust grains.  Because of variable conditions at the base of the wave over the past 10$^5$ years (indeed, the CND is a rather disturbed, non-equilibrated disk \cite{MS96}), such dust may not have been continuously available to highlight the wave.  This may also help explain why a similar nebula is not present on the opposite side of the CND.

An alternative scenario for understanding the Double Helix feature is that it be connected in some way with the linear radio filaments of the Galactic Center Radio Arc.  If the Northern extension of the Arc \cite{YZM88} is followed and extrapolated to Galactic latitudes beyond 0.5$^{\circ}$ (see fig 20b of \cite{YZHC04}), then it coincides approximately with the long axis of the Double Helix.  However, there is no continuous connection in the radio maps between the linear filaments and the Double Helix, and the only radio emission associated with the Double Helix lies outside the mid-IR strands (C. Law, personal communication).  There is so far no explanation for how a long bundle of linear, nonthermal filaments could culminate in helically wound, thermal structures.  Whether or not the CND hypothesis for the Double Helix is valid, further study of this feature should provide valuable insight into the Galactic center magnetic field.  

\begin{figure}[h]
\includegraphics[width=40pc]{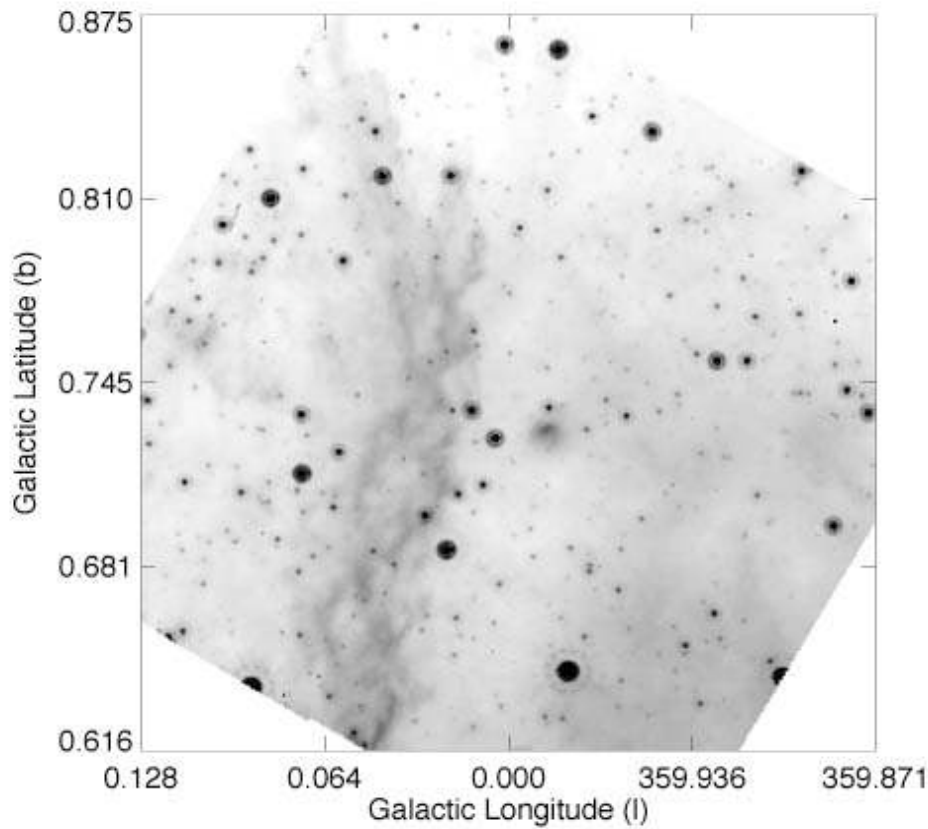}
\caption{\label{label2}The Double Helix Nebula, observed with the MIPS instrument on the Spitzer Space Telescope \cite{Morris06}.   This structure was also seen at shorter wavelengths, 3.6-8 $\mu$m, with the IRAC camera on Spitzer.}
\end{figure}

\section{Open Questions}

The questions that seem now to be the most compelling for guiding near-future research on the Galactic center magnetic field, besides those already mentioned above, are the following:

\begin{itemize}

\item Whether or not the central vertical field is more or less uniform, how and where does it merge with the azimuthal field of the Galactic disk?

\item If the Galactic center magnetosphere is defined as the region in which nonthermal radio filaments are observed, then its outer edge roughly coincides with the edge of the CMZ, with the Galaxy's inner inner Lindblad resonance, and with the transition from X1 to X2 gas orbits in the bar.  What is the interplay between these phenomena, at this critical juncture in the Galaxy?

\item Can high-resolution observations be used to obtain more detail on the points of interaction between cloud and intercloud fields?  This may best be done with a combination of radio and far-infrared polarization measurements.

\item What process produces the relativistic particles that illuminate the NTFs via their synchrotron emission?

\item At the moment, we lack consensus on the power source for the 10$^8$ K gas occupying much of the volume of the nuclear bulge.  Can we appeal to the stirring that takes place as clouds move through the field, leaving magnetosonic and Alfv\'en waves in their wake?  Or can the energy be supplied by magnetic field line annihilation of new vertical field constantly migrating inwards from the rest of the Galaxy?

\item What is the origin of the poloidal field?  Dynamo models have been hard-pressed to produce a dipole field like that observed, and a promising possibility is that the central field represents protogalactic field that has been concentrated over the history of the Galaxy by mass inflow \cite{Chandran00}.  Now is a propitious time to take these models to the next stage of sophistication.  

\end{itemize}

\ack
I gratefully acknowledge stimulating discussions with Steve Cowley.

\end{document}